\documentclass[%
 aip,
 amsmath,amssymb,
reprint, %onecolumn %%%%% Comment onecolumn to override one column format %%%%%%%
]{revtex4-1}

\usepackage{graphicx}% Include figure files
\usepackage{dcolumn}% Align table columns on decimal point
\usepackage{bm}% bold math
\usepackage[utf8]{inputenc}
\usepackage[T1]{fontenc}
\usepackage{mathptmx}
\usepackage{etoolbox}

\draft % marks overfull lines with a black rule on the right
\usepackage{subfigure}
\usepackage{multirow}
\usepackage{fancyhdr}
\usepackage{longtable}
\usepackage{parskip}

\usepackage{amsfonts}  % Common math fonts
\usepackage{amsmath}   % Common math functions
\usepackage{amssymb}   % Common math symbols
\usepackage{comment}   % to comment out large sections of text
\usepackage{xcolor}    % colored text
\usepackage{tabularx}  % table
\usepackage{array}     % fix column width table

\usepackage{url}
\usepackage{hyperref}

\makeatletter
\def\@email#1#2{%
 \endgroup
 \patchcmd{\titleblock@produce}
  {\frontmatter@RRAPformat}
  {\frontmatter@RRAPformat{\produce@RRAP{*#1\href{mailto:#2}{#2}}}\frontmatter@RRAPformat}
  {}{}
}%
\makeatother

\begin{document}
%\preprint{AIP/123-QED}
%\preprint{}

\title{Fundamental role of nonlocal orders in 1D Extended Bose-Hubbard Model} %Title of paper



\author{Nitya Cuzzuol}
\email[]{nitya.cuzzuol@polito.it}
\author{Arianna Montorsi}
\email[]{arianna.montorsi@polito.it}
%\thanks{}
%\affiliation {\it Department of Applied Science and Technology, Politecnico di Torino, Italy}
\affiliation {\it Institute for Condensed Matter Physics and Complex Systems, DISAT, Politecnico di Torino, I-10129 Torino, Italy}


\date{\today}

\begin{abstract}
Nonlocal order parameters capture the presence of correlated fluctuations between specific degrees of freedom, in otherwise disordered quantum matter. Here we provide a further example of their fundamental role, deriving the ground state phase diagram of the filling one extended Bose Hubbard model exclusively in terms of their ordering. By means of a density matrix renormalization group numerical analysis, we show that besides the (even) parity order characteristic of the Mott insulating phase, and the string order non vanishing in the Haldane insulator, the recently proposed odd parity order completes the picture, becoming nonvanishing at the transition from the normal superfluid to the paired superfluid phase. The above three nonlocal parameters capture all the distinct phases, including the density wave phase which local order is seen as the simultaneous presence of correlated fluctuations in different channels. They provide a unique tool for the experimental observation of the full phase diagram of strongly correlated quantum matter, by means of local density measurements.
\end{abstract}

 %, 75.10.Pq spin-chain model}% insert suggested PACS numbers in braces on next line

\maketitle %\maketitle must follow title, authors, abstract and \pacs

\begin{quotation}
The full phase diagram of the one dimensional three body constrained extended Bose Hubbard model is re-derived here in terms of solely three nonlocal order parameters, identified with the expectation values of appropriate disorder operators: the string operator, and the even and odd parity operators. Their finiteness unveils the emergence of distinct correlated density fluctuations for each disordered phase, which can persist at non-zero temperatures thanks to their nonlocal nature. These parameters can be detected by simple local density measurements in real experiments with quantum matter. 
\end{quotation}

\section{Introduction}
%P1: Intro: scientific motivation
The (1D) Hubbard model is the simplest paradigmatic model capturing the complex physics of strongly correlated particles (fermions or bosons) on a lattice. The model has been thoroughly studied in the last decades. In particular, at integer fillings and low temperatures, for repulsive on site density-density interaction $U$ a Mott insulating (MI) phase is observed, which escapes the Landau classification of spontaneous symmetry breaking (SSB) local orders \cite{Landau_1937, Landau_1980}. Indeed, it has been shown that the phase is characterized by a nonlocal order, in 1D known as charge parity order, in both the bosonic \cite{Berg_2008} and  the fermionic \cite{Montorsi_2012} cases. Whereas in 2D it is captured by its generalization, namely the brane parity order \cite{Fazzini_2017}. Despite their nonlocal nature, such orders are observed by local density measurements in atomic matter trapped onto optical lattices \cite{Endres_2011,Wei_2023,Hur_2024}. In fact, since they do not break any continuous symmetry, these order persist at sufficiently low temperatures.\\
In the 1D fermionic extended Hubbard model\cite{Sengupta_2002} the above MI phase is challenged by other insulating phases, which turn out to be associated to other types of nonlocal orders\cite{DallaTorre_2006, Barbiero_2013}, ultimately capturing the fundamental role of correlated quantum fluctuations in each of the possibly disordered distinct phases. The mathematical framework for nonlocal orders is provided by the concept of disorder operators\cite{Fradkin_2017}, which is not peculiar of insulating phases. Examples are spin parity order, observed in the Luther Emery liquid phase of 1D fermions\cite{Barbiero_2013}, or the brane spin parity, which is non vanishing in the superconducting phase of the 2D case\cite{Tocchio_2019}.  Most notably, besides parity orders, also string nonlocal orders appear in these systems, capturing symmetry protected topological (SPT) phases with non trivial topological features\cite{Gu_2009}. This is the case of the Haldane phase, which can be observed in insulating (HI) phases of bosons\cite{DallaTorre_2006} and fermions\cite{Barbiero_2013}, as well as in conducting phases\cite{Fazzini_2019, Montorsi_2020}.
%%%%%%%%%%%%%%%%%%%%Figure  1  begin %%%%%%%%%%%%%%%%%%%%%%%
\begin{figure*}
  \centering
  \includegraphics[width=\textwidth]{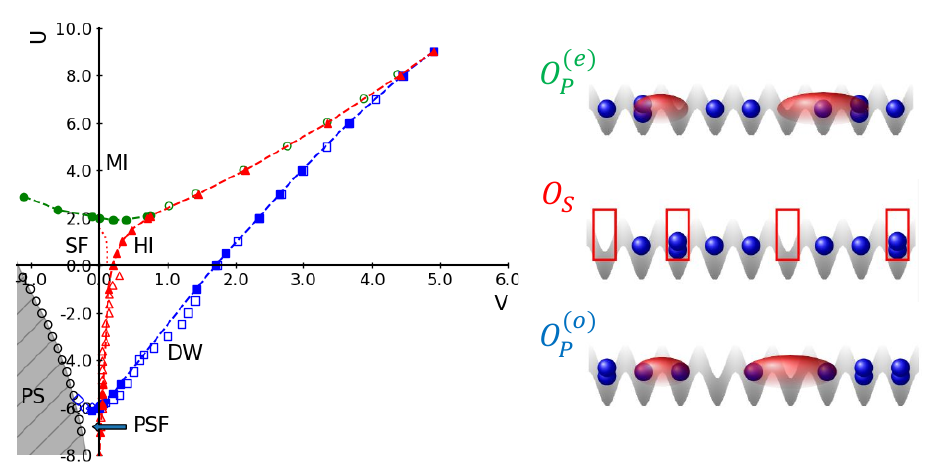}
\caption{\label{fig:EBHM1D3b_pd} 
%Start with a clause summarizing what is presented in the figure.
Phase diagram obtained by analyzing the nonlocal operators in eqs.(\ref{Op}),(\ref{Os}) for 3-body constrained 1D EBH model (left panel). Our results are represented by filled symbols, while the unfilled symbols correspond to the findings in \cite{Ejima_2014} for repulsive onsite interaction and \cite{Dalmonte_2011} for attractive onsite interaction. The full symbols of different colors signal that the expectation value of the corresponding nonlocal operator is different from zero (green for $O_P^{(e)}$, red for $O_S$, blue for $O_P^{(o)}$), separating phases with different nonlocal orders. The red dotted line close to the origin is the SF border predicted in \cite{Ejima_2014}.  The cartoons to the right schematically represent the correlated density fluctuations underlying the possible hidden nonlocal orders. The red ellipses highlight the two kind of pairs (holon-doublon or boson-boson) with finite correlation length which correspond to parity order (even or odd respectively). The empty rectangles highlight the alternation of holons and doublons in the background of single bosons characteristic of string order.
}
\end{figure*}
%%%%%%%%%%%%%%%%%%%%Figure 1 end %%%%%%%%%%%%%%%%%%%%%%%
On more general grounds -at least for correlated spinful fermions in one dimension \cite{Montorsi_2017} described in the low energy limit by decoupled sine Gordon models- the framework of SPT phases can be exploited to classify all distinct low temperature phases by means of appropriate nonlocal order parameters. In fact, taking advantage of such observation also for bosonic models, very recently it has been shown that another type of charge parity order (named odd charge parity) is finite in the paired superfluid (PSF) phase of the Bose-Hubbard model \cite{Cuzzuol_2024}.

Here we will explore the crucial role that nonlocal orders play in the full phase diagram of the Extended Bose Hubbard (EBH) model.  
Based on the theoretical framework, we will provide numerical evidence that the different possible correlated fluctuations shown in the right panel of Fig.(\ref{fig:EBHM1D3b_pd}) are in fact associated to three different nonlocal order parameters. Their finiteness uniquely identifies the disordered MI, HI, and PSF phases, as well as the locally ordered density wave (DW) phase. Each phase boundary is then identified by the vanishing of one of these operators, as shown in Fig.(\ref{fig:EBHM1D3b_pd}). There for completeness also the phase separated (PS) region is represented, not investigated here.

Our results, besides reinforcing evidence on the fundamental role of correlated fluctuations in disordered quantum matter, provide an immediate tool for the observation of these distinct phases by local density measurements in experiments at sufficiently low temperature.

\section{Model}
%description of the model, Hamiltonian and our knowledge of the phase diagram and phase transition
We will focus onto the study of the 1D Extended Bose-Hubbard model, that can be described by the Hamiltonian operator 
\begin{equation}\label{H_EBHM}
\begin{split}
    H =& - t \sum_{i} (b_i^{\dagger} b_{i+1} + b_{i+1}^{\dagger} b_{i})
    + \frac{U}{2} \sum_i n_i (n_i - 1) %\\
    %& - \mu \sum_{i} n_i\\
    %&
    + V \sum_{i} n_i n_{i+1} 
\end{split}
\end{equation}
where $b_i$, $b_i^\dagger$ are the bosonic creation and annihilation operators, with algebra $[b_i,b^\dagger_j]=\delta_{ij}$, $[b_i,b_j]=0=[b_i^\dagger,b_j^\dagger]$, $n_i = b^\dagger_i b_i$ is the number of particles at site $i$, $t$ is the hopping matrix element, and $U$ is the on-site interaction. % and $\mu$ is the chemical potential. 
In the following we will set $t=1$, and fix the average filling $\Bar{n}=1$. Moreover, we assume the three body constraint $(b_i^\dagger)^3=0$ to hold, which amounts to a truncation of the Hilbert space of each site to the three lowest occupation states $ 0 , 1 , 2$.

The model, besides capturing the SF-MI transition \cite{Fisher_1989} characteristic of these systems for $U\gtrsim 0$, gained a lot of attentions because for non vanishing nearest neighbor density-density interaction $V>0$ it also describes an HI phase, with non-trivial topological properties \cite{DallaTorre_2006, Berg_2008} and a finite nonlocal string order. 

For higher values of $V>0$ also a DW phase  appears\cite{Kuhner_2000, Rossini_2012}. For moderately attractive values of $V\lesssim 0$ instead, a SF phase is expected, which was observed to turn into a PSF phase in case of appropriate attractive $U$. The ground state phase diagram is reported in Fig.(\ref{fig:EBHM1D3b_pd}).

The low energy effective-field theory developed in the constrained case \cite{Berg_2008}, correctly captures the three insulating phases at filling $\Bar{n}=1$ in the repulsive interaction regime $U,V>0$, as well as a SF phase for weakly attractive $U$. The Hamiltonian $H$ is first mapped into a spin-$1$ model, then each spin-$1$ variable is written as a sum of two spin-$1/2$, which are mapped onto two spinless fermions via Jordan Wigner transforms. A standard bosonization analysis can be applied to the resulting fermionic model, which associates bosonic fields to the two fermionic degrees of freedom (DOF) and ends up in mapping the fermionic model into two sine Gordon Hamiltonians --decoupled in the symmetric and anti-symmetric combinations of the bosonic fields-- and a coupling term negligible in the low energy limit. We refer to \cite{Berg_2008} for the details of the calculation and of the subsequent renormalization group analysis. In fact, in case of fermions the above approach was exploited exhaustively to derive a general framework for associating nonlocal string and parity order parameters to each bosonic field\cite{Barbiero_2013, Montorsi_2017}. More  recently\cite{Cuzzuol_2024}, it was noticed that such procedure, when applied to the Bose Hubbard case, provides evidence of the transition to the PSF phase and of its characterization though a further nonlocal order parameter, named odd charge parity. Here we will prove that within the same theoretical framework, the full phase diagram of the EBH model can in fact be derived also at $V\geq 0$ through the appropriate nonlocal order parameters.

\section{Nonlocal Order Parameters}
The bosonization approach to constrained EBH model\cite{Berg_2008} associated to the symmetric (here even) bosonic field a local representations through the parity and string operators, which can be expressed in terms of local densities $n_i$. The MI and HI phases were identified with the pinning of such field to two possible distinct values, implying a finite %value for the 
expectation value of the even parity and string operators respectively. The previous mathematical property in fact signals the emergence of an hidden order, i.e. the ordering of a subset of DOF in the disordered background of the remaining DOF. In the present case of $3$ DOF per site, the different possible hidden orders are shown in the left panel of Fig.(\ref{fig:EBHM1D3b_pd}). The bosonization analysis\cite{Berg_2008,Cuzzuol_2024} proves that these correspond to finite expectation values of one or more of the following nonlocal operators:
\begin{equation}\label{Op}
O_P^{(\nu)}(j) = \prod_{i=0}^{j-1} \exp \left[ i\pi \delta_\nu n_i \right], \quad \nu=e,o
\end{equation}
\begin{equation}\label{Os}
O_S(j) = \delta n_{j} \prod_{i=0}^{j-1} \exp \left[ i\pi \delta n_i \right]
\end{equation}
known as parity ($O_P$) and string ($O_S$) operators. The further index $\nu$ in the parity operator, which can be even ($e$) or odd ($o$), was introduced in \cite{Cuzzuol_2024}, and corresponds to a different type of density fluctuation $\delta_\nu n_i$. Specifically,
\begin{equation}\label{deltan}
\begin{split}
    &\delta_e n_i = \delta n_i\doteq n_i - 1 \\
    &\delta_o n_i = n_i \quad .
\end{split}
\end{equation}
%explanation of the meaning of the NLOP
Let's discuss first the parity operator. It is evident %One can see 
from the definition of eq.(\ref{Op}) that the exponential factors assume either value $+1$ or $-1$ depending on the $\delta n_i$ on site i-th. Then the even parity operator $O_P^{(e)}$ will maintain on average a finite value only when the fluctuations with respect to the background of singly occupied sites, occur in the form of correlated nearby empty (holon) and doubly occupied (doublon) pairs. Reversely, for odd parity $O_P^{(o)}$ now the background amounts to disordered holons and doublons, and fluctuations must occur in pairs of nearby singly occupied sites to maintain its expectation value finite. This is shown in the right panel of Fig.(\ref{fig:EBHM1D3b_pd}).\\
Moreover, a non vanishing average value of the string operator $O_S$ in eq.(\ref{Os})  is obtained when holons and doublons are alternated though diluted in the disordered single site background. 
%%%%%%%%%%%%%%%%%%%%Figure  2  begin %%%%%%%%%%%%%%%%%%%%%%%
\begin{figure*}
  \centering
  \includegraphics[scale=1]{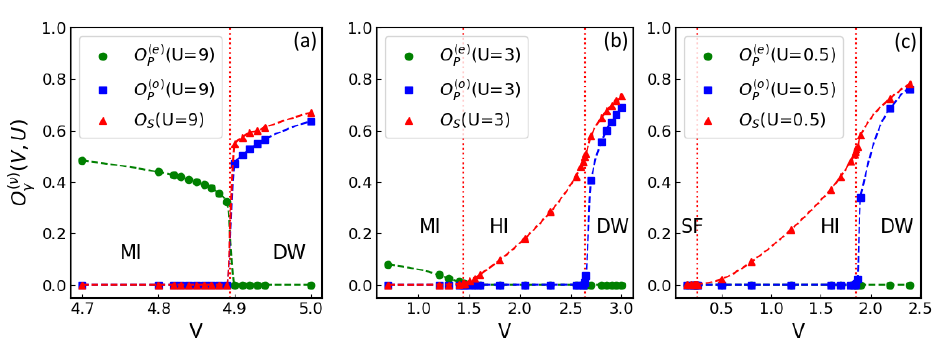}
  \caption{The analysis of the phase diagram depicted in Fig.(\ref{fig:EBHM1D3b_pd}) for $U>0$ and varying $V>0$ reveals four different possible phases: SF, MI, HI, DW. The three distinct scenarios are: (a) Plot of the case $U=9$, here there is a transition from a region dominated by even parity order (MI) to one where both string and odd parity are present (DW). (b) Plot of the case $U=3$, here between the two previous phases there is a region where only the string is finite (HI). (c) Plot of the case $U=0.5$, here there is first a region without any order (SF), then initially only the string is finite and finally also the odd parity becomes greater than zero.}
  \label{fig:posUcase}
\end{figure*}
%%%%%%%%%%%%%%%%%%%%Figure  2  end %%%%%%%%%%%%%%%%%%%%%%%
\\The results are summarized in Tab.(\ref{tab:OPtable}). Beyond the SF phase, four possible distinct phases can be obtained at filling $\Bar{n}=1$ from the ordering of the above nonlocal order parameters, according to the following table.  Phases having just one non vanishing nonlocal order (MI, HI, PSF) do not break any symmetry. The nonvanishing parameters become two in case of the DW phase: the simultaneous presence of two nonlocal orders ends up into the genuine local order of the DW phase, which breaks the translational symmetry.
\begin{table}[ht]
\begin{ruledtabular}
\begin{tabular}{cccccc}
NLOP        & SF& MI & HI & PSF & DW\\
\hline
$O_P^{(e)}$ & 0 & $\neq 0$ & 0 & 0 & 0\\
$O_S$ & 0 & 0 & $\neq 0$ & 0 & $\neq 0$\\
$O_P^{(o)}$ & 0 & 0 & 0 & $\neq 0$ & $\neq 0$\\
\end{tabular}
\caption{\label{tab:OPtable}%
Summary of the expectation values of the nonlocal order parameters (NLOP) in eqs.(\ref{Op})(\ref{Os}) for each phase according to bosonization analysis.}
\end{ruledtabular}
\end{table}
\\Also, in correspondence to each non vanishing nonlocal order parameter, similarly to what happens in the fermionic case, one or more excitation gaps are finite. In particular, we will be interested in
\begin{equation}\label{gap_eqs}
    \begin{split}
        & \Delta_n = E^{(1)}(N;L) - E^{(0)}(N;L)\\
        & \Delta_{\alpha = 1,2}(L) = E(N+\alpha; L) + E(N-\alpha;L) - 2E(N;L)\quad ,\\
    \end{split}
\end{equation}
namely the neutral gap, and the single and double particle excitation gaps for $N$ bosons on $L$ sites. These were already used to discuss the phase diagram at $U>0$\cite{Ejima_2014, Rossini_2012} and at $U<0$\cite{Dalmonte_2011}. As shall be clarified, %{\color{red}As shall be clarified,} %As we shall see, 
they behave differently in the different phases but unlike the nonlocal order parameters their finiteness cannot be used to identify uniquely all distinct phases.

\section{Results}
This section presents the outcomes of our analysis of the extended Bose-Hubbard model in one dimension (1D). As mentioned earlier, the comprehensive phase diagram of this model is here inferred by evaluating the expectation values of the nonlocal operators described in previous section. To accomplish this, we represented the state using the matrix product state (MPS) \cite{Fannes_1992, Klumper_1992, Ostlund_1995, Vidal_2003} formalism and applied the Density Matrix Renormalization Group (DMRG) algorithm to ascertain the ground state \cite{White_1992, White_1993, Ostlund_1995,Dukelsky_1998, Schollwock_2005, Schollwock_2011}.\\
In particular, we employed the infinite DMRG algorithm \cite{McCulloch_2008, tenpy} to approximate an infinite-size chain in our simulations. This approach enables us to circumvent significant issues related to boundary effects \cite{DegliEspostiBoschi_2016} when evaluating expectation of equations (\ref{Op}),(\ref{Os}). Moreover there is no need for extrapolation to the thermodynamic limit, requiring fewer DMRG states for obtaining the ground state; %Additionally, obtaining the ground state requires fewer DMRG states; 
we specifically set a maximal bond dimension of $\chi = 200$. %{\color{red}
This is enough to obtain a converged expectation value inside the phase characterized by the corresponding nonlocal order. The situation differs in computing the gaps, where the finite size algorithm is necessary.\\ %and open boundary conditions are preferable to reduce system degeneracy.
To derive our results, we imposed a maximum limit of 2 bosons per site to prevent the collapse of bosons onto a single site. While this restriction alters the model's delocalization of bosons in the lattice, the various phases identified in the literature for $U>0$ remain qualitatively consistent, as illustrated for larger maximal occupation numbers ($n_{\text{max}}=3$ in \cite{Rossini_2012} and $n_{\text{max}}=8$ in \cite{Deng_2013}). In fact, it has been demonstrated\cite{Kuhner_2000} that a maximum of $n_{\text{max}}=5$ provides a reasonable approximation of the system (in $U>0$ regime) without a specific limit on boson numbers.\\
The resulting phase diagram is presented in Fig.(\ref{fig:EBHM1D3b_pd}). Similar markers indicate equivalent transitions, with filled symbols representing our findings and empty symbols denoting results from the literature for the extended Bose-Hubbard model with the same three-body constraint. Besides a region of PS in which empty and doubly occupied regions coexists, five distinct phases emerge: MI, HI, DW, SF, PSF. Our transition points are identified where the expectation value of the operators become greater than zero (actually we assume the expectation is different from zero only when our estimate becomes greater then $3\times 10^{-3}$ for $j=200$ in eqs.(\ref{Op}),(\ref{Os})). While for the empty symbols we are referring to \cite{Ejima_2014} in the positive interaction regimes, and for negative onsite interaction the data are taken from \cite{Dalmonte_2011}.

\subsection{Repulsive U Regime}
%%%%%%%%%%%%%%%%%%%%Figure  3  begin %%%%%%%%%%%%%%%%%%%%%%%
\begin{figure*}[ht]
  \centering
  \includegraphics[scale=1]{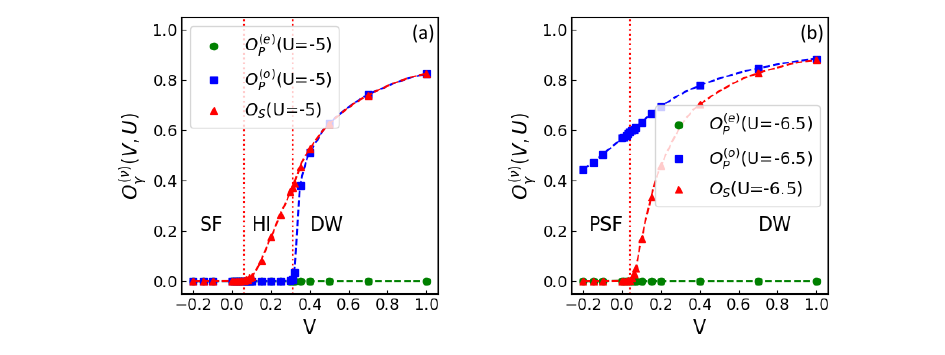}
  \caption{The analysis of the phase diagram depicted in Fig.(\ref{fig:EBHM1D3b_pd}) for $U<0$ and varying $V>0$ reveals four different possible phases: SF, PSF, HI, DW. The two distinct scenarios are: (a) Plot of the case $U=-5$, here there is an initial absence of order (SF). Then the first finite operator is the string (HI) and subsequently also the odd parity becomes non-zero (DW). (b) Plot of the case $U=-6.5$, here the odd parity is already greater than zero (PSF), and the transition from PSF to DW is signalled by $O_S$.}
  \label{fig:negUcase}
\end{figure*}
%%%%%%%%%%%%%%%%%%%%Figure  3  end %%%%%%%%%%%%%%%%%%%%%%%
In the region of $U>0$, it is possible to locate %{\color{red}it is possible to locate} %one can find
the following phases: SF, MI, HI, DW.\\
In order to show the different behaviours of the operators defined in eqs.(\ref{Op}) and (\ref{Os}), it is interesting to have a look at some sections of the phase diagram. In particular we will examine three different lines in the phase diagram, namely $U= 9, 3, 0.5$.\\
%For large onsite repulsion ($U=9$), by increasing $V$ from zero, one goes trough a discontinuous phase transition from the MI to DW. Indeed initially the dominant order is captured by the even parity, then after a critical $V$, one enters a region where both the odd parity and the string are greater than zero as presented in Fig.(\ref{fig:posUcase}.a)%first order
For large onsite repulsion ($U=9$), by increasing $V$ from zero, one goes, through a first order phase transition, from an initial phase where the dominant order is captured by the even parity (MI) to one where both the odd parity and the string are greater than zero (DW), as presented in Fig.(\ref{fig:posUcase})(a)%first order
.\\
When the onsite interaction is reduced, for instance to $U=3$ as in Fig.(\ref{fig:posUcase})(b), at sufficiently weak nearest-neighbor interaction the even parity is the only operator different from zero: this is the MI phase. In this scenario, by further increasing the nearest-neighbor interaction the system first enters --through a Gaussian type phase transition-- a phase where the string operator is finite (HI), and then, upon an Ising type phase transition, also the odd parity becomes non vanishing in the DW phase.\\
Further decreasing $U$ towards the origin, for instance in the case $U=0.5$ shown in Fig.(\ref{fig:posUcase})(c),  $O_P^{(e)}$ becomes vanishing. At weak $V$ there is neither local or nonlocal order and the system is in a normal SF phase. By increasing $V$ the string parameter slowly becomes nonvanishing at the SF-HI transition. Then, by further increasing $V$, also the odd parity becomes different from zero and we are again in the DW phase. Even if both transition are continuous they belong to different universality classes. The SF-HI is a BKT transition while the HI-DW transition is an Ising one.\\
%comparison with Ejima
In the phase diagram Fig.(\ref{fig:EBHM1D3b_pd}), we compared our findings with those reported in \cite{Ejima_2014} by determining the central charge \cite{Giamarchi_2003}. %Their approach involved extracting phase transitions through an analysis of entanglement properties. Specifically, they determined the central charge by evaluating the von Neumann entropy. %Based on the predicted universality class and central charge, the MI to HI transition occurs when the system becomes critical for $c = 1$, and the HI to Density Wave DW transition occurs for $c=1/2$. 
%{\color{green} Cosa vuoi dire? However, their use of the central charge method precludes the study of the SF-HI phase transitions, since SF can be interpreted as a Luttinger liquid with $c = 1$ \cite{Berg_2008, Giamarchi_2003}.}\\
%Apart from this aspect, 
Their results closely align with ours, and the two phase diagrams exhibit a significant overlap.\\
We also investigated the behavior of the gaps defined in equation (\ref{gap_eqs}) in this region. The results are summarized in the subsequent Tab.(\ref{tab:gap}), for the repulsive $U$ region as well.
\begin{table}[ht]
\begin{ruledtabular}
\begin{tabular}{cccccc}
Gap        & SF& PSF & MI & HI  & DW\\
\hline
$\Delta_n$ & 0 & 0 & $\neq 0$ & $\neq 0$  & $\neq 0$\\
$\Delta_1$ & 0 & $\neq 0$ & $\neq 0$ & $\neq 0$  & $\neq 0$\\
$\Delta_2$ & 0 & 0 & $\neq 0$ & $\neq 0$  & $\neq 0$\\
\end{tabular}
\caption{\label{tab:gap}%
Gaps introduced in eq.(\ref{gap_eqs}) for each phase. As explained in the text, the opening (or temporary vanishing) of the gaps supports the results obtained by the nonlocal order parameters. A null gap is intended in the thermodynamic limit.}
\end{ruledtabular}
\end{table}
Since the SF phase is the only one where all gaps are zero, both SF-MI and SF-HI are signaled by any of the three gaps becoming finite. The scenario differs for other phase transitions in the region of $U>0$, since all MI, HI, DW phases possess a finite gap. Indeed the transition between the distinct MI and HI phases is signaled by a temporary vanishing of both the charge ($\Delta_1$ and $\Delta_2$) and the neutral ($\Delta_n$) gaps, since the two phases are protected by different symmetries. %However, in both phases the three gaps are non-vanishing, as well as in the nearby DW phase, so that they cannot be used to distinguish them. 
While the transition to the latter is signaled solely by the vanishing of the neutral gap.
\\In the repulsive $U$ region of the phase diagram the gap $\Delta_2$ provides no additional information.

\subsection{Attractive U Regime}
Switching to the attractive interaction regime $U<0$, it is possible to find %{\color{red}it is possible to find} %one can find 
the following phases: SF, PSF, HI, DW and the PS region.\\
As before let's have a look to the different behaviours of the three nonlocal operators defined in eqs.(\ref{Op})(\ref{Os}). Contrarily to the case of $U>0$, now the $O_P^{(e)}$ will be always zero and the correspondent order is absent. Thus the phase diagram can be discussed here through the behavior of $O_P^{(o)}$ and $O_S$ solely. We can consider two representative lines of the phase diagram, $U= -5, -6.5$, which are located above and below the transition line from SF to PSF phase. These are shown in Fig.(\ref{fig:negUcase}). 
\\In particular, in (a) we consider $U=-5$: there, increasing $V$ from zero, the system initially remains in the phase where all nonlocal orders are zero (SF). Subsequently, similarly to the weakly repulsive $U$ regime, first the string $O_S$ becomes different from zero (HI) and then also the odd parity (DW).\\% The operators defining the phases are the same as before for $U>0$.\\
Moving down to stronger onsite attraction between bosons, in Fig.(\ref{fig:negUcase})(b) we consider the case $U=-6.5$. There also at weakly attractive $V$ a PSF phase characterised by non-null odd parity is present. Contrary to the SF phase, this phase now is partly gapped and characterized by the emergence of the hidden order identified in \cite{Cuzzuol_2024}. By further increasing $V$, one enters directly the region in which both string and odd parity orders are finite (DW).\\% Also in this case the rising of the string operator is exponentially slow, representative of the BKT transition.\\
%spiegare come hanno ottenuto i loro diagrammi di fase
Also for the $U<0$ region we can compare our findings with existing results \cite{Dalmonte_2011}, obtaining a good agreement.\\
Moreover, also in this region we investigated the behavior of the different gaps defined in eq.(\ref{gap_eqs}). The results were anticipated in Tab.(\ref{tab:gap}). For sufficiently weakly attractive $U$, the scenario is the same as for weakly repulsive $U$, in agreement with the phase diagram of Fig.(\ref{fig:EBHM1D3b_pd}). Entering the PSF phase instead, just the single particle gap opens, to signal the cost in energy of adding an unpaired single particle fluctuation\cite{Cuzzuol_2024}. On the contrary, the neutral gap remains zero (for an infinite chain), since the PSF phase is still superfluid. So that the transition from PSF to DW is in fact signaled by $\Delta_n$ becoming different from zero. The same happens if instead one moves from PSF to PS. 

\section{Conclusions}
In summary, we presented an alternative numerical derivation of the full phase diagram of the extended Bose Hubbard model in 1D, which is based solely on the behavior of three nonlocal order parameters. These amount, besides the string and (even) parity disorder operators introduced in \cite{Berg_2008} to describe the MI and HI insulator phases, also to the odd parity operator introduced in \cite{Cuzzuol_2024} to describe the PSF phase. The latter captures the correlated fluctuations of pairs of single particles in a disordered background of holons and doublons. Besides describing the PSF phase, odd parity is here found to be non vanishing also in presence of a finite string order, describing the alternation of holons and doublons. In this case, a true local order appears, and the system enters a SSB DW phase.\\
The phase diagram obtained by the interplay of the three above disorder operators is shown in Fig.(\ref{fig:EBHM1D3b_pd}) and is in full accordance with previous results in literature \cite{Ejima_2014, Dalmonte_2011}. Thus our analysis is capable of characterizing distinct conducting and insulating phases of these systems by the appropriate order or disorder operator, at the same time identifying all the transition lines and the type of transition.\\
Notably, since the presence of nonlocal orders does not violate Mermin Wagner theorem, the MI, HI, and PSF nonlocal orders could persist even at non zero temperature, and can thus be observed in experiments with quantum matter. Another significant advantage of the identified nonlocal orders discussed here is that their measurement can be gained by just local density measures. Indeed, in case of the MI phase, the (even) parity order was already measured in this way both in 1D and in 2D\cite{Endres_2011,Wei_2023,Hur_2024}. We expect that a similar result could be obtained also in case of the odd parity characteristic of the PSF phase, which in principle can be generalized also to 2D as a odd brane parity order\cite{Fazzini_2017}.

%P21: Acknowledgements
\section*{ACKNOWLEDGMENTS}
AM sincerely thanks David Campbell for introducing her several years ago to the rich physics of the extended Hubbard model and for stimulating her with his profound insights to delve into its hidden orders.\\
We offer our thanks for the financial support from the ICSC – Centro Nazionale di Ricerca in High Performance Computing, Big Data and Quantum Computing, funded by European Union – NextGenerationEU (Grant number CN00000013). Computational resources were provided by HPC@POLITO (http://www.hpc.polito.it). Calculations were performed using the TeNPy Library (version 0.10.0)\cite{tenpy}.
\section*{Author Declaration}
\subsection*{Data Availability Statement}
The data that support the findings of this study are available from the corresponding author upon reasonable request.
\subsection*{Conflict of Interest Statement}
The authors have no conflicts to disclose.
\subsection*{Author Contributions}
\textbf{Nitya}: Formal Analysis, Data Curation, Writing – original draft, review and editing\\
\textbf{Arianna}: Funding Acquisition, Conceptualization, Supervision, Writing – review and editing

%\begin{center}
%\line(1,0){70}
%\end{center}
%\newpage
%\section*{References}
%\bibliographystyle{apsrev}
\bibliography{myRef.bib}

\end{document}